\documentclass[ prd, twocolumn, twoside, preprintnumbers, superscriptaddress, nofootinbib, natbib, bm, amsrefs ]{revtex4-2}
\usepackage{graphicx}% Include figure files
\usepackage{xcolor}
\usepackage{amsmath, amstext, amsfonts,amssymb, amsthm}
\usepackage{comment}
\usepackage[utf8]{inputenc}
\usepackage{multirow}

\definecolor{BlueViolet}{rgb}{0.2, 0.00, 0.7}
\definecolor{Blue}{rgb}{0.15, 0.00, 0.9}
\definecolor{lightblue}{rgb}{0.15, 0.35, 0.95}
\definecolor{kitgreen}{rgb}{0,
0.58823 %150/255,
0.50980 %130/255
}
\usepackage[%dvipdfmx,
colorlinks=true, linkcolor=lightblue,citecolor=lightblue,urlcolor=kitgreen]{hyperref}

\newcommand{\Eprint}[1]{\href{#1}}

\definecolor{lb}{rgb}{.74,.83,.9}
\definecolor{ly}{rgb}{1,.92,.8}
\definecolor{lr}{rgb}{.98,.85,.87}

\begin{document}
\title{Endpoint anatomy of the baryon--meson sum rule in $b\to s\nu\bar\nu$}
\author{Syuhei Iguro}
\affiliation{Institute for Advanced Research, Nagoya University, Nagoya 464-8601, Japan}
\affiliation{Kobayashi-Maskawa Institute for the Origin of Particles and the Universe, Nagoya University, Nagoya 464--8602, Japan}

%%% Please do not use your own command in abstract for the arXiv submission %%%
\begin{abstract}
Baryon--meson sum rules enable cross-checks among exclusive mesonic and baryonic measurements.
Unlike their heavy-to-heavy counterparts, analogous heavy-to-light relations lack an explicit heavy-quark-symmetry origin.
We investigate their origin in the $b\to s\nu\bar\nu$ system by studying the relevant helicity amplitudes at the two kinematic endpoints. 
Although the sum rule follows algebraically from the simple short-distance structure, its coefficient depends nontrivially on the hadronic dynamics.
At maximum recoil, the coefficient for the daughter baryons ${\cal B}=\Lambda,\Xi$ is consistent with the approximate value $1/2$ within current form-factor uncertainties.
At zero recoil, by contrast, a kinematic helicity projection enforces its exact vanishing. 
The fully integrated coefficients are numerically close to $1/4$.
However, unlike the one-quarter coefficient in the heavy-to-heavy system, they are not local symmetry coefficients but integrated quantities shaped by the behavior between the two endpoints.
This understanding may guide the construction of baryon--meson sum rules in more general heavy-to-light decays.\\ 
---------------------------------------------------------------------------------------------------------------------------------
\end{abstract}
\maketitle

%%%%%%%%%%%%%%%%%%%%%%%%%%%%%%%%%%%%%
\section{Introduction}
\label{sec:intro}
%%%%%%%%%%%%%%%%%%%%%%%%%%%%%%%%%%%%%
Baryon--meson sum rules relate observables in exclusive decays with different hadronic final states.
They allow a baryonic observable to be predicted from mesonic measurements, or vice versa, with reduced sensitivity to the underlying short-distance scenario \cite{Blanke:2018yud,Blanke:2019qrx,Fedele:2022iib,Duan:2024ayo}.
Such relations therefore provide useful consistency tests among measurements that probe the same quark-level transition through different spin and helicity structures.

The established example is the semileptonic $b\to c\ell\bar\nu$ sum rule relating $B\to D^{(*)}\ell\bar\nu$ and $\Lambda_b\to\Lambda_c\ell\bar\nu$.
Originally motivated by the long-standing tensions in $R_{D^{(*)}}$ \cite{HFLAV24,Iguro:2024hyk}, it provides a largely new-physics (NP) model-independent consistency test among mesonic and baryonic measurements.
It can be written schematically as \cite{Endo:2025lvy}
\begin{align}
 \frac{R_{\Lambda_c}}{R_{\Lambda_c}^{\rm SM}}\simeq
 \alpha_{\Lambda_c}\frac{R_D}{R_D^{\rm SM}}+\left(1-\alpha_{\Lambda_c}\right)\frac{R_{D^*}}{R_{D^*}^{\rm SM}} .
\label{eq:h2h_sum_rule}
\end{align}
Heavy quark symmetry \cite{Isgur:1989vq,Isgur:1990yhj,Isgur:1990pm,Neubert:1993mb} guarantees the underlying differential relation in the heavy-quark limit \cite{Endo:2025fke,Endo:2025lvy}.
At zero recoil, the exact $\alpha_{\Lambda_c}=1/4$ admits a particularly transparent Clebsch--Gordan interpretation in terms of the pseudoscalar and vector members of the heavy-meson spin doublet \cite{Endo:2025set}.
The framework has since been extended to angular observables, to $SU(3)_{\rm F}$-rotated decay modes and to excited decay modes and so on, broadening the range of experimental cross-checks \cite{Endo:2025cvu,Iguro:2026xgi,Endo:2026afe,Endo:2026dxr}.
These developments naturally raise the question of whether analogous relations can be understood beyond heavy-to-heavy transitions.

Numerically successful baryon--meson relations have also been pointed out in heavy-to-light transitions \cite{Duan:2024ayo,Lee:2026jvq,Kitahara:2026doj,Endo:2026blh}.
Examples include the charged-current $b\to u\ell\bar\nu$ modes $B\to\pi\ell\bar\nu$, $B\to\rho\ell\bar\nu$, and $\Lambda_b\to p\ell\bar\nu$, as well as the rare $b\to s\nu\bar\nu$ modes $B\to K^{(*)}\nu\bar\nu$ and $\Lambda_b\to\Lambda\nu\bar\nu$.
Remarkably, the coefficients of the integrated relations are often found numerically close to $1/4$, familiar from the heavy-to-heavy system.
No heavy-quark symmetry, however, relates the light final-state hadrons.
It is therefore unclear whether this numerical similarity reflects a common physical mechanism or merely a coincidence.
Studying the analytical structure may clarify the origin of this similarity and may also guide the construction of analogous relations in other semileptonic processes.

A complementary clue may come from kinematic endpoint relations.
At zero recoil, the restoration of spatial rotational symmetry reduces the number of independent helicity amplitudes and enforces model-independent relations among them.
These endpoint relations \cite{Zwicky:2013eda} have been developed for both mesonic and baryonic decays, and their common and distinctive features have also been briefly discussed \cite{Hiller:2013cza,Hiller:2021zth}.
Their implications for the hadronic coefficient of a baryon--meson sum rule, however, have not been explored.
In particular, it remains unclear whether the endpoint helicity structure can explain how the coefficient of the baryon--meson relation varies across phase space and why its integrated value is numerically close to $1/4$.

The simple $b\to s\nu\bar\nu$ system provides a particularly clean setting for addressing these questions.
Experimental interest in this transition has also been reinforced by the first evidence for $B^+\to K^+\nu\bar\nu$ reported by Belle II \cite{Belle-II:2023esi}.
Unlike charged-lepton modes such as $b\to s\ell\bar\ell$, it is free from photon-pole effects and long-distance charmonium contributions.
Moreover, within the minimal operator basis considered here, all normalized observables depend on only two short-distance invariants.
This allows us to separate the algebraic origin of the baryon--meson relation from the hadronic dynamics that determine its coefficient.
Our main focus is therefore the kinematic behavior of this coefficient at maximum and zero recoil.
The numerical analysis illustrates how the coefficient varies between the two endpoints using current form-factor inputs.

The remainder of this paper is organized as follows.
In Sec.~\ref{sec:setup}, we derive the differential baryon--meson relation and clarify its algebraic origin.
In Sec.~\ref{sec:endpoint}, we investigate its endpoint structure and present a numerical illustration.
Section~\ref{sec:conc} contains our conclusions and outlook.

%%%%%%%%%%%%%%%%%%%%%%%%%%%%%%%%%%%%%
\section{Baryon--meson relation}
\label{sec:setup}
%%%%%%%%%%%%%%%%%%%%%%%%%%%%%%%%%%%%%
We consider the $b\to s\nu\bar\nu$ transition within the effective Hamiltonian,
\begin{align}
{\cal H}_{\rm eff} &= -\frac{4G_F}{\sqrt{2}}V_{tb}V_{ts}^* \frac{\alpha}{4\pi}C_L^{\rm SM} \times \nonumber\\
&~~~~~~~~\sum_{i,j} \left[ \left(\delta_{ij}+C_L^{ij}\right)O_L^{ij} + C_R^{ij}O_R^{ij} \right] +{\rm h.c.},
\label{eq:Heff}
\end{align}
where 
\begin{align}
O_{L,R}^{ij} = \left(\bar s\gamma_\mu P_{L,R}b\right) \left(\bar\nu_i\gamma^\mu(1-\gamma_5)\nu_j\right), 
\label{eq:operators}
\end{align}
and $C_L^{\rm SM}=-X_t/\sin^2\theta_W$. 
Here, $X_t$ includes higher-order corrections \cite{Buchalla:1993bv,Misiak:1999yg,Buchalla:1998ba,Brod:2010hi}, while $C_L^{ij}$ and $C_R^{ij}$ parameterize NP contributions and vanish in the Standard Model (SM) \cite{Buras:2014fpa,Allwicher:2023xba}.
The chiral projectors are defined as $P_{L,R}=(1\mp\gamma_5)/2$ and $i,j=1,2,3$ label the neutrino flavors.
We neglect neutrino masses and assume that right-handed neutrinos are decoupled.
The overall normalization, including $C_L^{\rm SM}$, cancels from all normalized observables considered below.

The hadronic vector and axial-vector currents probe the combinations
\begin{align}
 C_+^{ij} =\delta_{ij}+C_L^{ij}+C_R^{ij},\,\,\,\,
 C_-^{ij}=\delta_{ij}+C_L^{ij}-C_R^{ij},
\label{eq:Cpm}
\end{align}
respectively, where the SM contribution is included in $\delta_{ij}$.
After summing over the unobserved neutrino flavors, all short-distance dependence can be collected into two real combinations,
\begin{align}
 S_+=\frac{1}{3}\sum_{i,j}|C_+^{ij}|^2,\,\,\,\,
 S_-=\frac{1}{3}\sum_{i,j}|C_-^{ij}|^2.
\label{eq:Spm}
\end{align}
They are normalized such that $S_+=S_-=1$ in the SM.
The quantities $S_+$ and $S_-$ are not helicity amplitudes, but the short-distance strengths multiplying the hadronic vector and axial-vector responses.

To compare equivalent positions in the phase space of different channels, we introduce $q^2=(p_\nu+p_{\bar\nu})^2$, the invariant mass squared of the neutrino pair, and define
\begin{align}
 x=\frac{q^2}{q^2_{{\rm max},X}},\,\,\,\,
 q^2_{{\rm max},X}=\left(M-m\right)^2,\,\,\,
 0\leq x\leq1,
\label{eq:x}
\end{align}
where $X$ labels the decay channel, while $M$ and $m$ denote the masses of the corresponding parent and daughter hadrons, respectively.
For each mode, we define the normalized differential rate
\begin{align}
 \mu_X(x)=
 \left(\frac{d\Gamma_X}{dx}\right)\biggl/
 \left(\frac{d\Gamma_X}{dx}\right)_{\rm SM}.
\label{eq:local_mu}
\end{align}
The three classes of observables can then be written as
\begin{align}
 \mu_K(x)&=S_+,\nonumber\\
 \mu_{K^*}(x) &= a_{K^*}(x)S_+ + \left[1-a_{K^*}(x)\right]S_-,\nonumber\\
 \mu_{\cal{B}}(x) &= b_{\cal{B}}(x)S_+ + \left[1-b_{\cal{B}}(x)\right]S_-, \,\,\,\, {\cal B}=\Lambda,\Xi .
\label{eq:local_observables}
\end{align}
The pseudoscalar mode is purely vector-current-like, whereas the vector-meson and baryonic modes probe both short-distance directions.
We refer to $a_{K^*}(x)$ and $b_{\cal B}(x)$ as the mesonic and baryonic weights, respectively.
These weights encode the channel-dependent hadronic effects.

Since all three normalized rates depend only on $S_+$ and $S_-$, the baryonic rate can be written as a linear combination of the two mesonic rates, 
\begin{align}
\mu_{\cal B}(x) = \alpha_{\cal B}(x)\mu_K(x) + \left[1-\alpha_{\cal B}(x)\right]\mu_{K^*}(x),
\label{eq:differential_sum_rule}
\end{align}
where
\begin{align}
\alpha_{\cal B}(x) = \frac{b_{\cal B}(x)-a_{K^*}(x)} {1-a_{K^*}(x)}.
\label{eq:local_alpha} 
\end{align} 
We refer to $\alpha_{\cal B}(x)$ as the baryon--meson interpolation coefficient.
It specifies the position of the baryonic response between the $K$-like and $K^*$-like short-distance responses.
For instance, $\alpha_{\cal B}(x)=1$ corresponds to a purely $K$-like response, whereas $\alpha_{\cal B}(x)=0$ corresponds to a purely $K^*$-like response.

Equation~(\ref{eq:differential_sum_rule}) is exact within the minimal operator basis.
Its existence follows from the two-dimensional short-distance structure of Eq.\,(\ref{eq:local_observables}) and does not require a dynamical meson--baryon symmetry.
The nontrivial hadronic information is instead encoded in $\alpha_{\cal B}(x)$ through the phase-space-dependent hadronic weights $a_{K^*}(x)$ and $b_{\cal B}(x)$.
We examine this quantity at the large- and zero-recoil endpoints below.

%%%%%%%%%%%%%%%%%%%%%%%%%%%%%%%%%%%%%
\section{Endpoint relations}
\label{sec:endpoint}
%%%%%%%%%%%%%%%%%%%%%%%%%%%%%%%%%%%%%
We first consider the large-recoil endpoint, $x=0$, corresponding to $q^2=0$ in each channel.
For the $K^*$ mode, the local hadronic weight is written as 
\begin{align}
 a_{K^*}(q^2) = \frac{W_V(q^2)} {W_V(q^2)+W_{A_1}(q^2)+W_{A_{12}}(q^2)},
\label{eq:aKstar_local}
\end{align}
where $W_V$, $W_{A_1}$, and $W_{A_{12}}$ denote the corresponding contributions to the differential rate, including the relevant phase-space and helicity factors.
Explicit decay rate expressions and conventions necessary to calculate $W$ are provided in Appendix~\ref{app:meson}.
The first two terms describe the transverse vector and axial-vector responses, respectively, while $W_{A_{12}}$ describes the longitudinal axial-vector response.
Near $q^2=0$, the transverse kernels vanish, whereas the longitudinal axial-vector kernel approaches a finite, nonzero limit: 
\begin{align} 
 W_V&(q^2)= \mathcal{O}(q^2),\,\,W_{A_1}(q^2) = \mathcal{O}(q^2),\nonumber\\  
 &W_{A_{12}}(q^2) = W_{A_{12}}(0)+\mathcal{O}(q^2),
\label{eq:Kstar_large_recoil_scaling} 
\end{align} 
with $W_{A_{12}}(0)\neq0$.
It therefore follows that $a_{K^*}(0)=0$.
This does not imply that the transverse vector-current amplitudes vanish at the amplitude level.
Rather, their contributions to the physical rate are suppressed by the overall $q^2$ factor. 
The longitudinal axial-vector amplitude contains the compensating helicity normalization factor $1/\sqrt{q^2}$ and remains finite in the rate. 

For a baryonic channel, the vector and axial-vector kernels take the schematic form 
\begin{align}
 K_V^{\cal{B}}(q^2) &\propto q^2\sqrt{\lambda_{\cal{B}}}\,Q_- \left[ \frac{(M+m)^2}{q^2}f_+^{\cal{B}}(q^2)^2 +2f_\perp^{\cal{B}}(q^2)^2 \right], \nonumber\\
 K_A^{\cal{B}}(q^2) &\propto q^2\sqrt{\lambda_{\cal{B}}}\,Q_+ \left[ \frac{(M-m)^2}{q^2}g_+^{\cal{B}}(q^2)^2 +2g_\perp^{\cal{B}}(q^2)^2 \right], 
\label{eq:baryon_kernels} 
\end{align}
where $Q_\pm=(M\pm m)^2-q^2$ and $\lambda_{\cal{B}}(M^2,\,m^2,\,q^2)=Q_+Q_-$ are introduced.
Appendix~\ref{app:baryon} summarizes the relevant decay rate relations.
At $q^2=0$, the explicit $q^2$ factor is canceled by the longitudinal terms proportional to $1/q^2$, while the transverse terms vanish.
The remaining kinematic factors satisfy 
\begin{align}
 Q_-(0)(M+m)^2 = Q_+(0)(M-m)^2 = (M^2-m^2)^2 . 
\end{align}
Using $a_{K^*}(0)=0$, the interpolation coefficient at maximum recoil is entirely determined by the relative vector and axial-vector baryonic responses as 
\begin{align} 
\alpha_{\cal{B}}(0) = b_{\cal{B}}(0) = \frac{f_+^{\cal{B}}(0)^2} {f_+^{\cal{B}}(0)^2+g_+^{\cal{B}}(0)^2}. 
\label{eq:alpha_large_recoil}
\end{align}

At large recoil, the energetic light quark is approximately collinear, and the vector and axial-vector helicity form factors are governed by the same leading soft dynamics \cite{Wang:2011uv,Feldmann:2011xf,Mannel:2011xg}.
This motivates the approximate relation $f_+^{\cal{B}}(0)\sim g_+^{\cal{B}}(0)$ and hence 
\begin{align} 
 \alpha_{\cal{B}}(0)\simeq\frac{1}{2}. 
\label{eq:large_recoil_half} 
\end{align}
Unlike the zero-recoil relation discussed below, Eq.~(\ref{eq:large_recoil_half}) is not an exact kinematic identity.
Its accuracy is controlled by perturbative, power-suppressed, and light-quark-mass corrections to the large-recoil form-factor relation.

We next consider the opposite endpoint, $x=1$, corresponding to the channel-specific zero-recoil limit $q^2=q^2_{\rm max}$.
In the broader low-recoil region, mesonic $B\to K^*\ell^+\ell^-$ amplitudes admit a simplified description based on a local operator product expansion (OPE) and heavy-quark form-factor relations \cite{Bobeth:2012vn}.
For the baryonic mode $\Lambda_b\to\Lambda\ell^+\ell^-$, analogous simplifications follow from heavy-quark form-factor relations at low recoil \cite{Boer:2014kda}.
At the exact kinematic endpoint, additional relations arise from the restoration of rotational symmetry as the recoil momentum vanishes \cite{Zwicky:2013eda,Hiller:2013cza,Hiller:2021zth}.
These endpoint relations are kinematic rather than consequences of the low-recoil OPE.
They apply directly to the hadronic helicity amplitudes relevant for the di-neutrino modes considered here.

For the modes of interest, the endpoint behavior suppresses the vector-current response relative to the axial-vector response.
Consequently, both the $K^*$ and baryonic modes become aligned with the same short-distance direction, $S_-$, at zero recoil.
We now demonstrate this behavior for the two classes of final states.

For the $K^*$ mode, the relevant kernels scale as
\begin{align}
 W_V(q^2) &\propto \lambda_{K^*}^{3/2}(q^2), \nonumber\\
 W_{A_1}(q^2)&,\,W_{A_{12}}(q^2) \propto \lambda_{K^*}^{1/2}(q^2), 
 \label{eq:Kstar_zero_recoil_scaling}
\end{align}
where $\lambda_{K^*}(q^2)=\lambda(m_B^2,m_{K^*}^2,q^2)$.
Because $\lambda_{K^*}^{1/2}$ is proportional to the recoil momentum and vanishes at $q^2=q^2_{\rm max}$, the vector-current kernel carries additional powers of the recoil momentum relative to the axial-vector kernels.
It therefore vanishes more rapidly at the endpoint, and the local vector-current weight satisfies
\begin{align}
 a_{K^*}(1)=0.
\end{align}
Thus, the normalized $K^*$ rate becomes purely $S_-$-like at zero recoil.

This suppression can also be understood directly from the antisymmetric momentum structure of the $P\to V$ vector-current matrix
element,
\begin{align}
 \langle V(k,\varepsilon)|\bar s\gamma^\mu b|P(p)\rangle
 \propto \epsilon^{\mu\nu\rho\sigma}\varepsilon_\nu^*p_\rho k_\sigma.
 \label{eq:PV_vector_structure}
\end{align}
At zero recoil, the parent and daughter momenta are parallel, so this antisymmetric structure vanishes.
The axial-vector current, by contrast, can couple directly to the vector-meson polarization.

For the baryonic modes, zero recoil corresponds to
$Q_-\to 0$ and $Q_+\to 4Mm$.
Equation~\eqref{eq:baryon_kernels} then gives
\begin{align}
 \frac{K_V^{\cal B}(q^2)}{K_A^{\cal B}(q^2)}
 \propto \frac{Q_-}{Q_+} \to 0,
\end{align}
provided that the endpoint form factors remain regular.
The baryonic vector-current weight therefore approaches
\begin{align}
 b_{\cal B}(1)=0.
\end{align}
Thus, the normalized baryonic rate also becomes purely $S_-$-like at zero recoil.
Together with $a_{K^*}(1)=0$, Eq.~\eqref{eq:local_alpha} immediately gives
\begin{align}
 \alpha_{\cal B}(1)=0.
 \label{eq:alpha_zero_recoil}
\end{align}
Equivalently, the baryonic and $K^*$ observables have the same short-distance response at the endpoint, so no $K$-like component is required in the baryon--meson relation.

The current-level interpretation of the baryonic suppression is also simple.
At zero recoil, the spatial vector current is proportional to the recoil momentum and is therefore suppressed.
By contrast, the spatial axial-vector current reduces to a spin operator and remains non-vanishing relative to the common phase-space suppression.
The temporal vector current itself need not vanish.
At the endpoint, however, it is parallel to $q^\mu$ and drops out after contraction with the conserved massless-neutrino current, $q_\mu j_\nu^\mu=0$.
The physical baryonic response is therefore aligned with the $S_-$ direction.

This common $S_-$ alignment of the $K^*$ and baryonic modes is not a generic consequence of large $q^2$.
The $B\to K$ mode has no final-state spin polarization. 
Its axial-vector matrix element vanishes, and the mode remains purely $S_+$-like throughout phase space.
The zero-recoil relation $\alpha_{\cal B}(1)=0$ instead follows from the helicity projection of the spinful vector-meson and baryonic final states.

%%%%%%%%%%%%%%%%%%%%%%%%%%%%%%%%%%%%%%%%%%%
\begin{figure}[t] 
\centering 
\includegraphics[width=0.9\columnwidth]{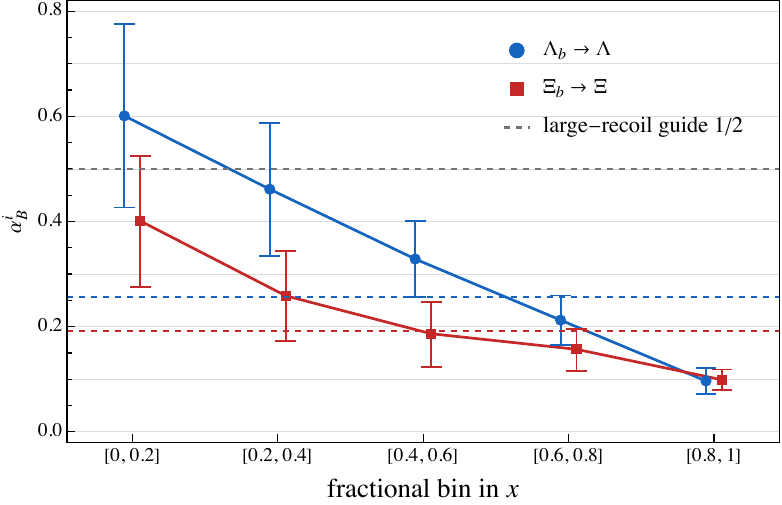} 
\caption{
Fractional-bin baryon--meson interpolation coefficients evaluated in the five equally divided bins defined in Appendix~\ref{app:bin}. 
The blue filled circles and red squares correspond to $\Lambda_b\to\Lambda$ and $\Xi_b\to\Xi$, respectively, with the associated error bars.
The uncertainty prescriptions are detailed in the main text.
The colored dashed lines indicate the fully integrated central results, which lie close to $1/4$.
The gray dashed line at $1/2$ is a large-recoil reference.
}
\label{fig:alpha_MC} 
\end{figure}
%%%%%%%%%%%%%%%%%%%%%%%%%%%%%%%%%%%%%%%%%%%

%%%%%%%%%%%%%%%%%%%%%%%%%%%%%%%%%%%%%%%%%%%
\begin{figure*}[t]
\centering 
\includegraphics[width=0.45\linewidth] {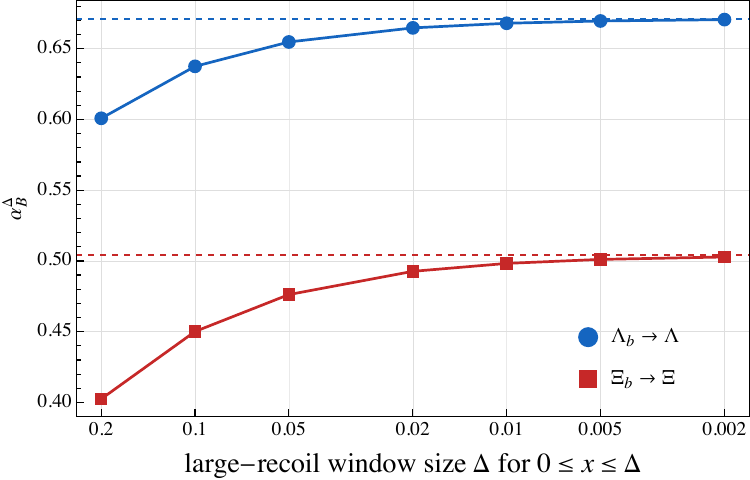} 
\,\,\,\,\,\,\,
\includegraphics[width=0.45 \linewidth] {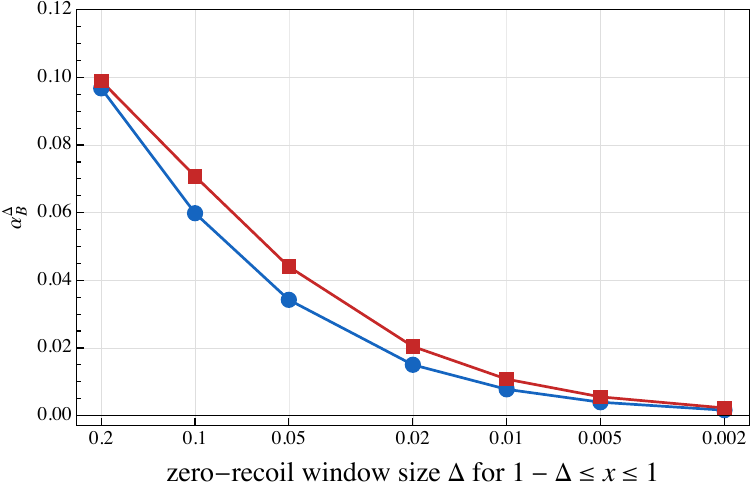}
\caption{
Convergence of the window-integrated interpolation coefficients toward the two kinematic endpoints.
The blue circles and red squares correspond to $\Lambda_b\to\Lambda$ and $\Xi_b\to\Xi$, respectively.
For the $\Lambda_b\to\Lambda$ trajectory, the points are evaluated from the nominal central form-factor parameters, while for $\Xi_b\to\Xi$, the points are the medians of the correlated Monte Carlo (MC) distributions.
The left and right panels show the nested windows $x\in[0,\Delta]$ and $x\in[1-\Delta,1]$, respectively.
The dashed lines in the left panel denote the endpoint values obtained from the corresponding central prescriptions.
}
\label{fig:endpoint}
\end{figure*}
%%%%%%%%%%%%%%%%%%%%%%%%%%%%%%%%%%%%%%%

To illustrate how the interpolation coefficient connects the two endpoint regimes, we now evaluate it using current form-factor determinations. 
We use the $B\to K^*$ form factors of Ref.\,\cite{Gao:2024vql}, the lattice-QCD $\Lambda_b\to\Lambda$ form factors of Ref.\,\cite{Detmold:2016pkz}, and the $\Xi_b\to\Xi$ form factors of Ref.\,\cite{Farrell:2026swf}.\footnote{
In implementing the $B\to K^*$ input, the quantity denoted by $\mathcal A_1-\mathcal A_2$ in the supplementary fit information of Ref.\,\cite{Gao:2024vql} is used to reconstruct the conventional $A_2$ form factor and subsequently the helicity form factor $A_{12}$ entering Eq.~\eqref{eq:app_Kstar_kernels}.
Here, $\mathcal A_1$ and $\mathcal A_2$ denote the rescaled form factors used in Ref.\,\cite{Gao:2024vql}, and should not be confused with the conventional $A_1$, $A_2$, or $A_{12}$.
This implementation reproduces the longitudinal-polarization prediction of Ref.\,\cite{Gao:2024vql} and provides an independent normalization check.
}
The form-factor uncertainties are propagated using the channel-specific prescriptions described in Appendix~\ref{app:uncertainty}.
The $B\to K^*$ and baryonic form-factor inputs are treated as statistically independent.

For an integration region $\Omega$ defined in Appendix~\ref{app:bin}, the interpolation coefficient is
constructed as
\begin{equation}
 \alpha_{\cal B}^{\Omega}
 =\frac{b_{\cal B}^{\Omega}-a_{K^*}^{\Omega}}
 {1-a_{K^*}^{\Omega}}.
\label{eq:alpha_full}
\end{equation}
For $\Lambda_b\to\Lambda$, the nominal central values and the total nominal-higher-order uncertainties are quoted according to the prescription described in Appendix~\ref{app:uncertainty}.
For the $\Xi_b\to\Xi$ results quoted below, the central values and uncertainties refer to the MC means and sample standard deviations, unless explicitly stated otherwise.

For the individual fully integrated hadronic fractions, we obtain
\begin{align}
 a_{K^*}^{\rm full} &= 0.21 \pm 0.02, \nonumber\\
 b_\Lambda^{\rm full} &= 0.41 \pm 0.04, \nonumber\\
 b_\Xi^{\rm full} &= 0.36 \pm 0.03.
\label{eq:integrated_hadronic_fractions}
\end{align}
Propagating the mesonic and baryonic form-factor uncertainties gives
\begin{align}
 \mu_\Lambda^{\rm full}
 &= (0.26\pm0.05)\,\mu_K^{\rm full}
 +(0.74\mp0.05)\,\mu_{K^*}^{\rm full}, \nonumber\\
 \mu_\Xi^{\rm full}
 &= (0.19\pm0.05)\,\mu_K^{\rm full}
 +(0.81\mp0.05)\,\mu_{K^*}^{\rm full}.
\end{align}
The two coefficients in each relation are fully anti-correlated because their sum is fixed to unity.

The maximum-recoil and fully integrated results can be summarized as
\begin{align}
 \alpha_\Lambda(0) &= 0.67\pm0.19,
 &\alpha_\Lambda^{\rm full} &= 0.26\pm0.05, \nonumber\\
 \alpha_\Xi(0) &= 0.50\pm0.16,
 &\alpha_\Xi^{\rm full} &= 0.19\pm0.05,
\label{eq:numerical_endpoint_summary}
\end{align}
whereas the exact zero-recoil relation is $\alpha_{\cal B}(x)|_{x\to1}=0$.

The $\Xi_b\to\Xi$ channel gives a smaller fully integrated interpolation coefficient, with a comparable absolute precision.
The precision of its form-factor description benefits from the dispersive bounds together with the endpoint and asymptotic constraints. 
Consequently, the fully integrated $\Xi_b\to\Xi$ rate is more closely aligned with the $K^*$-like short-distance direction than the $\Lambda_b\to\Lambda$ rate.
Although symmetry-breaking corrections can shift the central values away from $1/2$, both maximum-recoil results remain compatible with the approximate large-recoil relation $\alpha_{\cal B}(0)\simeq1/2$ within current uncertainties.
The agreement is particularly close for the central $\Xi_b\to\Xi$ result, whereas the nominal $\Lambda_b\to\Lambda$ fit prefers a somewhat larger value.

Figure~\ref{fig:alpha_MC} shows the coefficients in the five equal fractional bins defined in Appendix~\ref{app:bin}.
For $\Lambda_b\to\Lambda$, the marker positions are evaluated with the nominal form-factor fit of Ref.\,\cite{Detmold:2016pkz}, while the error bars show the total nominal--higher-order uncertainties defined in Appendix~\ref{app:uncertainty}.
For $\Xi_b\to\Xi$, the markers and error bars denote the medians and central $68\%$ intervals of the correlated MC distributions.
The correlated $B\to K^*$ form-factor uncertainty is included in both channels.
The colored dashed lines correspond to the fully integrated central results.
The gray dashed one at $1/2$ is a large-recoil reference.
For the uncertainty, we apply the nominal--higher-order prescription of the underlying form-factor analysis \cite{Detmold:2016pkz} directly to the normalized integrated coefficient, retaining the correlations between its numerator and denominator.
The integrated $\Lambda_b\to\Lambda$ relation was previously evaluated in Ref.\,\cite{Kitahara:2026doj}, and our result agrees with their latest analysis.

For both baryonic channels, the finite-bin coefficient decreases from the large-recoil region toward zero recoil.
The first bin reflects the approach toward the large-recoil regime, where the approximate value $1/2$ provides a useful reference.
The last bin exhibits the onset of the exact zero-recoil alignment. 
The fully integrated coefficients lie between these limits and should therefore be interpreted as rate-integrated hadronic coordinates, not as local endpoint values.
The proximity of $\alpha_\Lambda^{\rm full}$ to $1/4$ neither implies that $\alpha_\Lambda(x)$ is approximately constant over phase space nor signals a local symmetry coefficient analogous to that in the heavy-to-heavy system.
Rather, it reflects the integrated effect of the trajectory between the two endpoint regimes.

Figure~\ref{fig:endpoint} makes the approach to the two endpoint limits explicit using nested integration windows.
For the initial windows $x\in[0,\Delta]$, the coefficients approach the finite maximum-recoil central values, $\alpha_\Lambda^{\rm central}(0)\simeq0.67$ and $\alpha_\Xi^{\rm central}(0)\simeq0.50$.
The central $\Xi_b\to\Xi$ value lies closer to $1/2$ than the nominal $\Lambda_b\to\Lambda$ value.
Both channels remain compatible with this expectation within their current uncertainties.
More precise form-factor determinations would be required to resolve channel-dependent deviations.

For the terminal windows $x\in[1-\Delta,1]$, both coefficients decrease toward zero as the window is reduced. 
Unlike the maximum-recoil values, this common zero-recoil limit is fixed by the helicity endpoint relations.
The two panels therefore exhibit the interpolation between a form-factor-dependent maximum-recoil limit and a kinematically fixed $K^*$-aligned zero-recoil limit.
The endpoint limits themselves should not, however, be interpreted as representative values for the finite bins.
Because the differential rates are phase-space suppressed at the exact endpoints, the first and last bins receive substantial contributions from regions away from $x=0$ and $x=1$, respectively.
Their values therefore characterize the responses integrated over finite kinematic intervals rather than the corresponding local endpoint limits.
The same consideration applies more broadly to the fully integrated coefficient, which is shaped by the rate-weighted evolution over the entire phase space and is not represented by either endpoint value.

%%%%%%%%%%%%%%%%%%%%%%%%%%%%%%%%%%%%%
\section{Conclusion and discussion}
\label{sec:conc}
%%%%%%%%%%%%%%%%%%%%%%%%%%%%%%%%%%%%%
We have clarified the origin and endpoint structure of the baryon--meson sum rule in the minimal $b\to s\nu\bar\nu$ system.
Such sum rules provide short-distance-model-independent consistency tests among mesonic and baryonic measurements that probe the same quark-level transition through different hadronic structures.
The relation itself follows algebraically because the normalized rates depend on only two independent short-distance invariants, and therefore does not require a dynamical meson--baryon symmetry.
The nontrivial hadronic information is instead encoded in the phase-space dependence of the interpolation coefficient $\alpha_{\cal{B}}(x)$.

The two endpoints reveal distinct physical mechanisms.
At maximum recoil, the approximate form-factor relation $f_+^{\cal{B}}(0)\simeq g_+^{\cal{B}}(0)$ leads to $\alpha_{\cal{B}}(0)\simeq1/2$, up to symmetry-breaking corrections.
At zero recoil, by contrast, a kinematic helicity projection enforces the exact limit $\alpha_{\cal{B}}(1)=0$.
The integrated value $\alpha_\Lambda^{\rm full}\simeq1/4$ is therefore not a local symmetry coefficient analogous to that in the heavy-to-heavy system. 
Instead, it is an integrated hadronic coordinate shaped by the trajectory between the two endpoint regimes.
The numerical similarity of the heavy-to-light and heavy-to-heavy coefficients thus has a different physical origin.
The numerical results based on current form-factor determinations illustrate this endpoint-controlled trajectory and support this interpretation within present uncertainties.
Future high-statistics flavor facilities, including Belle II and proposed Tera-$Z$ factories, could probe the endpoint-controlled trajectory through complementary finite-bin measurements \cite{Belle-II:2018jsg,Amhis:2023mpj,Ai:2024nmn}.

Separating the algebraic origin of the sum rule from the hadronic origin of its coefficient provides a useful framework for constructing further consistency tests in heavy-to-light decays.
Analogous endpoint analyses could help to understand the discovered approximate baryon--meson relations in $b\to u\tau\bar\nu$ and $b\to s\mu\bar\mu$ transitions, where lepton-mass effects, a larger operator basis, and nonlocal contributions must be taken into account \cite{Iguro:WIP}.

%%%%%%%%%%%%%%%%%%%%%%%%%%%%%%%%%%%%%%%%%%%%%%%%%%%%%%%
\section*{Acknowledgements}
%%%%%%%%%%%%%%%%%%%%%%%%%%%%%%%%%%%%%%%%%%%%%%%%%%%%%%%
The author thanks Teppei Kitahara, Kota Sasaki, Tim Kretz, Satoshi Mishima, Motoi Endo, Ryoutaro Watanabe, and Hiroyasu Yonaha for helpful discussions and comments on the numerical analysis.
%---------------------------------------------------------------------------
This work is supported by JSPS KAKENHI Grant Numbers 22K21347, 24K23939, and 25K17385 and the Toyoaki scholarship foundation.
%---------------------------------------------------------------------------

%%%%%%%%%%%%%%%%%%%%%%%%%%%%%%%%%%%%%%%%%%%%%%%%%%%%%%%% 
\appendix
%%%%%%%%%%%%%%%%%%%%%%%%%%%%%%%%%%%%%
\section{Decay rate and numerical procedure} 
%%%%%%%%%%%%%%%%%%%%%%%%%%%%%%%%%%%%%
In this Appendix, we collect the differential decay rate expressions for mesonic and baryonic modes in Appendix~\ref{app:meson} and Appendix~\ref{app:baryon}, respectively.
The definitions of the integrated coefficients used in the numerical analysis are given in Appendix~\ref{app:bin}.
For the form-factor parameterizations, fit coefficients, and covariance matrices, we refer to Refs.\,\cite{Gao:2024vql,Detmold:2016pkz,Farrell:2026swf}.
The propagation of the form-factor uncertainties is described in Appendix~\ref{app:uncertainty}.

%%%%%%%%%%%%%%%%%%%%%%%%%%%%%%%%%%%%%%%%%%%
\subsection{$B\to K^{(*)}\nu\bar\nu$} 
\label{app:meson} 
%%%%%%%%%%%%%%%%%%%%%%%%%%%%%%%%%%%%%%%%%%%
We use the kinematic quantities and short-distance invariants defined in the main text. 
All $q^2$-dependent phase-space and helicity factors are included in the kernels defined below.
The factors ${\cal N}_X$ therefore contain only $q^2$-independent normalization constants and cancel from the normalized hadronic weights.

We first collect the differential-rate expressions for the mesonic modes.
For $B\to K\nu\bar\nu$, only the vector quark current contributes, and the differential rate can be written as
\begin{align}
\frac{d\Gamma_K}{dq^2} = {\cal N}_K\, W_K(q^2)\,S_+, 
\label{eq:app_K_rate} 
\end{align} 
where 
$W_K(q^2) = \lambda_K^{3/2}(q^2)\, f_+^K(q^2)^2$ and the arguments of the K\"all\'en function are abbreviated.
The precise definition of the common normalization factor ${\cal N}_K$ is immaterial for the normalized observables considered in the main text. 
Equation~\eqref{eq:app_K_rate} directly gives $\mu_K(x)=S_+$. 
For $B\to K^*\nu\bar\nu$, the differential rate is decomposed into vector, transverse-axial, and longitudinal-axial contributions: 

\begin{align} 
\frac{d\Gamma_{K^*}}{dq^2} = {\cal N}_{K^*} \bigl[ W_V(q^2)S_+ + \{W_{A_1}(q^2)+W_{A_{12}}(q^2)\}S_- \bigr].
\label{eq:app_Kstar_rate} 
\end{align}
In the form-factor convention used in this work, the kernels are 

\begin{align} 
W_V(q^2) &= \frac{2q^2\lambda_{K^*}^{3/2}(q^2)} {(m_B+m_{K^*})^2} V(q^2)^2, \\
W_{A_1}(q^2) &= 2q^2\lambda_{K^*}^{1/2}(q^2) (m_B+m_{K^*})^2 A_1(q^2)^2, \\ 
W_{A_{12}}(q^2) &= 64m_B^2m_{K^*}^2 \lambda_{K^*}^{1/2}(q^2) A_{12}(q^2)^2. 
\label{eq:app_Kstar_kernels}
\end{align}
The longitudinal form factor is defined as 
\begin{widetext}
\begin{align} 
A_{12}(q^2) = \frac{ (m_B+m_{K^*})^2 (m_B^2-m_{K^*}^2-q^2)A_1(q^2) - \lambda_{K^*}(q^2)A_2(q^2) }{ 16m_Bm_{K^*}^2(m_B+m_{K^*}) }. 
\label{eq:app_A12} 
\end{align}
\end{widetext}
The local vector-current weight is therefore 
\begin{align}
a_{K^*}(q^2) = \frac{W_V(q^2)} {W_V(q^2)+W_{A_1}(q^2)+W_{A_{12}}(q^2)}. 
\label{eq:app_aKstar}
\end{align} 
These expressions reproduce the large- and zero-recoil scalings used in Sec.~\ref{sec:endpoint}.

%%%%%%%%%%%%%%%%%%%%%%%%%%%%%%%%%%%%%%%%%%%
\subsection{Baryonic modes} 
\label{app:baryon}
%%%%%%%%%%%%%%%%%%%%%%%%%%%%%%%%%%%%%%%%%%%
For ${\cal B}=\Lambda,\Xi$, the differential rate can be expressed as 
\begin{align} 
\frac{d\Gamma_{\cal{B}}}{dq^2} = {\cal N}_{\cal{B}} \left[ K_V^{\cal{B}}(q^2)S_+ + K_A^{\cal{B}}(q^2)S_- \right], 
\label{eq:app_baryon_rate}
\end{align} 
where 
\begin{align} 
K_V^{\cal{B}}(q^2) &= 
q^2\sqrt{\lambda_{\cal{B}}}\,Q_- \left[ \frac{(M+m)^2}{q^2}f_+^{\cal{B}}(q^2)^2 + 2f_\perp^{\cal{B}}(q^2)^2 \right],\nonumber\\
K_A^{\cal{B}}(q^2) &= 
q^2\sqrt{\lambda_{\cal{B}}}\,Q_+ \left[ \frac{(M-m)^2}{q^2}g_+^{\cal{B}}(q^2)^2 + 2g_\perp^{\cal{B}}(q^2)^2 \right].
\label{eq:app_baryon_kernels} 
\end{align} 
Here, $f_+^{\cal B}$ and $g_+^{\cal B}$ describe the longitudinal vector and axial-vector helicity responses, respectively, while $f_\perp^{\cal B}$ and $g_\perp^{\cal B}$ describe the corresponding transverse responses.
The corresponding local baryonic weight is 
\begin{align} b_{\cal{B}}(q^2) = \frac{K_V^{\cal{B}}(q^2)} {K_V^{\cal{B}}(q^2)+K_A^{\cal{B}}(q^2)}. 
\label{eq:app_baryon_weight} 
\end{align} 
The endpoint limits of Eq.~\eqref{eq:app_baryon_weight} reproduce the maximum- and zero-recoil relations derived in the main text.

%%%%%%%%%%%%%%%%%%%%%%%%%%%%%%%%%%%%%%%%%%%
\subsection{Binned and endpoint-window coefficients}
\label{app:bin}
%%%%%%%%%%%%%%%%%%%%%%%%%%%%%%%%%%%%%%%%%%%
For an integration region $\Omega\subset[0,1]$ in the normalized variable $x$, we define the integrated hadronic weights by 
\begin{align} 
a_{K^*}^{\Omega} &= 
\frac{ \displaystyle\int_\Omega dx\,W_V(x) }{ \displaystyle\int_\Omega dx\, [W_V(x)+W_{A_1}(x)+W_{A_{12}}(x)] }, \nonumber\\ 
b_{\cal{B}}^\Omega &= 
\frac{ \displaystyle\int_\Omega dx\,K_V^{\cal{B}}(x) }{ \displaystyle\int_\Omega dx\, [K_V^{\cal{B}}(x)+K_A^{\cal{B}}(x)] },~~~
\alpha_{\cal{B}}^\Omega = \frac{b_{\cal{B}}^\Omega-a_{K^*}^\Omega} {1-a_{K^*}^\Omega}. 
\label{eq:app_integrated_coefficients}
\end{align} 
Upon changing variables from $q^2$ to $x=q^2/q_{{\rm max},X}^2$, the constant Jacobian $dq^2=q_{{\rm max},X}^2\,dx$ cancels between the numerator and denominator of each integrated hadronic weight.
In each integral, the kernels are evaluated at the channel-specific momentum transfer $q^2=x\,q_{{\rm max},X}^2$. 
The vector and axial-vector responses are integrated separately before constructing $\alpha_{\cal{B}}^\Omega$. 
Consequently, $\alpha_{\cal{B}}^\Omega$ is not an unweighted average of the local coefficient $\alpha_{\cal{B}}(x)$.

The fully integrated coefficient corresponds to $\Omega_{\rm full}=[0,1]$. 
The five equal fractional bins used in Fig.~\ref{fig:alpha_MC} are 
\begin{align} 
\Omega_i = \left[ \frac{i-1}{5}, \frac{i}{5} \right], \qquad i=1,\ldots,5. 
\label{eq:app_five_bins} 
\end{align}
For the endpoint convergence test in Fig.~\ref{fig:endpoint}, we use the nested windows 
\begin{align}
\Omega_{\rm max}(\Delta) = [0,\Delta], \,\,\,\, \Omega_{\rm zero}(\Delta) = [1-\Delta,1],
\label{eq:endpoint_windows} 
\end{align}
with $\Delta = 0.2,\, 0.1,\, 0.05,\, 0.02,\, 0.01,\, 0.005,\, 0.002$.
For smooth form factors, the corresponding small-window behavior takes the form 
\begin{align} 
\alpha_{\cal{B}}^{[0,\Delta]} = \alpha_{\cal{B}}(0)+{\cal O}(\Delta), \,\,\,\, \alpha_{\cal{B}}^{[1-\Delta,1]} = {\cal O}(\Delta), 
\label{eq:app_window_scaling} 
\end{align} 
consistent with the numerical convergence shown in Fig.~\ref{fig:endpoint}.

%%%%%%%%%%%%%%%%%%%%%%%%%%%%%%%%%%%%%%%%%%% 
\subsection{Propagation of form-factor uncertainties} 
\label{app:uncertainty} 
%%%%%%%%%%%%%%%%%%%%%%%%%%%%%%%%%%%%%%%%%%% 
The form-factor uncertainties are propagated using channel-specific prescriptions.
For $B\to K^*$ and $\Xi_b\to\Xi$, we generate correlated samples of the corresponding form-factor coefficients from their multivariate distributions.
For each sample, the vector and axial-vector contributions are integrated separately over the region $\Omega$. 
The integrated weights $a_{K^*}^{\Omega}$ and $b_{\Xi}^{\Omega}$ are then constructed before evaluating the nonlinear ratio according to Eq.~\eqref{eq:alpha_full}.
The $B\to K^*$ and $\Xi_b\to\Xi$ form-factor inputs are sampled independently.
For numerical results quoted in the text, we use the MC mean and sample standard deviation, while the markers and error bars in Fig.~\ref{fig:alpha_MC} denote the median and central $68\%$ interval, respectively. 

For $\Lambda_b\to\Lambda$, we follow the uncertainty prescription of Ref.\,\cite{Detmold:2016pkz}. 
For a generic observable, the nominal fit yields the central value $O_{\rm nom}$ and statistical uncertainty $\sigma_{O,\rm nom}$. 
The same observable is evaluated using the higher-order fit, yielding $O_{\rm HO}$ and $\sigma_{O,\rm HO}$. 
The systematic uncertainty is defined as \cite{Detmold:2016pkz}
\begin{align}
\sigma_{O,\mathrm{syst}} &= \max\!\Bigg[ \left|O_{\rm HO}-O_{\rm nom}\right|, \nonumber\\[-2pt] &\,\,\,\,\,\,\,\,\,\,\,\,
\Theta\!\left(\sigma_{O,\rm HO}-\sigma_{O,\rm nom}\right) \sqrt{ \sigma_{O,\rm HO}^2-\sigma_{O,\rm nom}^2 } \Bigg], 
\label{eq:DM_systematic_prescription} 
\end{align}
where $\Theta$ is the Heaviside step function.
The total uncertainty is 
\begin{align}
\sigma_{O,\mathrm{tot}} = \sqrt{ \sigma_{O,\rm nom}^2+\sigma_{O,\mathrm{syst}}^2 }. 
\label{eq:DM_total_uncertainty} 
\end{align}

We apply this prescription directly at the observable level to $b_\Lambda^\Omega$ and $\alpha_\Lambda^\Omega$ for the full phase-space region, the five fractional bins, and the endpoint windows.
The numerator and denominator of each normalized hadronic weight are evaluated using the same form-factor parameter set and covariance matrix, so that their correlations are retained.

When applying this prescription to $\alpha_\Lambda^\Omega$, the same $B\to K^*$ form-factor determination is used in both the nominal and higher-order $\Lambda_b\to\Lambda$ evaluations. 
The $B\to K^*$ and $\Lambda_b\to\Lambda$ form-factor determinations are treated as statistically independent.
The central values shown for $\Lambda_b\to\Lambda$ are evaluated with the nominal fit, while the error bars in Fig.~\ref{fig:alpha_MC} represent the symmetric total uncertainties obtained from Eqs.~\eqref{eq:DM_systematic_prescription} and \eqref{eq:DM_total_uncertainty}.

%%%%%%%%%%%%%%%%%%%%%%%%%%%%%%%%%%%%%%%%%%%%%%%%%%%
%%%%%%%%%%%%%%%%%%%%%%%%%%%%%%%%%%%%%%%%%%%%%%%%%%%
\bibliographystyle{utphys28mod}
\bibliography{refs}
\end{document}